\documentclass[reprint,aps,prl,amsmath,amssymb,superscriptaddress, preprintnumbers]{revtex4-2}

\usepackage[utf8]{inputenc}
\usepackage[T1]{fontenc}
\usepackage{amsmath}
\usepackage{graphicx}
\usepackage{appendix}
\usepackage{adjustbox}
\usepackage{bm}
\usepackage{epstopdf} 
\usepackage[colorlinks=true, allcolors=blue]{hyperref}
\usepackage{orcidlink}
\usepackage{multirow}
\usepackage{booktabs}
\newcommand{\nc}{\newcommand}       
\nc{\no}[1]{:\!\! #1 \!\!:}

\newcommand{\gscas}{\affiliation{Graduate School of China Academy of Engineering Physics, Beijing 100193, China}}

\begin{document}







\title{Cutoff-independent predictions from nuclear lattice effective field theory}

\author{Chen-Can Wang}
\gscas
\author{Jia-Ai Shi}
\gscas
\author{Bing-Nan Lu}
\gscas
\email{bnlv@gscaep.ac.cn}

\date{\today}

\begin{abstract} 

Cutoff independence is an essential requirement for the predictive power of nuclear \textit{ab initio} calculations based on effective field theory (EFT). 
While it is conventionally assumed that such invariance necessitates high-order interactions and complex many-body forces, we present a minimal chiral nuclear force that exhibits remarkable cutoff independence across a broad range from light to medium-mass nuclei and sub-saturated nuclear matter.
Our framework comprises only contact terms up to next-to-leading order, a single three-nucleon contact force, and a leading-order one-pion-exchange potential, all constrained strictly in the $A \leq 3$ sector. 
Despite its simplicity, this interaction accurately reproduces experimental binding energies up to $^{40}\text{Ca}$ with unexpectedly small residual cutoff dependencies of only a few MeV. 
We demonstrate that the use of a lattice-inspired \emph{absolute}-momentum regulator efficiently suppresses high-momentum modes, resolving the overbinding problem for soft chiral forces without invoking complex many-body forces. 
These results establish a robust and economic foundation for EFT-based \textit{ab initio} calculations in both continuum and lattice frameworks.

\end{abstract}
\maketitle

\textit{Introduction.}---Recent decades have witnessed remarkable progress in nuclear \textit{ab initio} calculations, driven by advances in both high-fidelity nuclear interactions and sophisticated quantum many-body methods~\cite{machleidt2023fbs,papenbrock2024arxiv,navratil2016ps}. Chiral effective field theory (EFT) has established itself as the leading framework for describing nuclear forces, organizing nuclear forces into series of contact and pion-exchange terms regulated with certain momentum cutoffs~\cite{weinberg1990plb,epelbaum2009rmp,machleidt2011pr,hammer2020rmp}. 
The flexibility in regularization prescriptions, power counting schemes and optimization strategies inspires a variety of chiral force models~\cite{entem2003prc,epelbaum2005npa,ekstroem2013prl,epelbaum2015epja,reinert2018epja}. 

Renormalization group (RG) invariance is a cornerstone of nuclear EFT, demanding that physical observables remain independent of the regulator cutoff~\cite{machleidt2020epja, thomas2025arxiv, furnstahl2021fbs}. 
Analytical studies have primarily focused on the two-nucleon scattering~\cite{kaplan1998plb, nogga2005prc, epelbaum2017npb, kvinikhidze2018epja, epelbaum2018epja, epelbaum2020epja, contessi2024plb, yang2025prc}.
While investigations in three-~\cite{bedaque1999prl, wang2012prl,epelbaum2017epja, song2017prc}, 
and four-body systems~\cite{platter2005plb,  kirscher2013plb, klein2018epja, avraham2026plb, shi2026plb}  
have demonstrated that a single three-body force are sufficient for renormalizing these systems,
achieving similar RG-invariance for medium-mass nuclei and nuclear matter remains a formidable numerical challenge.
Current \textit{ab initio} frameworks typically circumvent this issue by relying on fixed cutoffs or empirical refitting to heavier systems~\cite{ekstroem2013prl, shirokov2016plb, logoteta2016plb,drischler2019prl, hoppe2019prc, soma2020prc, huther2020plb, jiang2024prc}, often involving complex 3NF parameterizations to improve phenomenological descriptions. Although recent efforts have begun exploring the cutoff dependence of medium-mass systems~\cite{schiavilla2021prc,yang2021prc,sanchez2023,contessi2025arxiv}, the findings remain inconclusive and sensitive to the chosen optimization scheme. Consequently, establishing a cutoff-independent chiral EFT beyond the lightest nuclei is still a critical open problem essential for advancing the predictive power of \textit{ab initio} nuclear theory.

A central challenge is identifying a minimal EFT interaction that simultaneously achieves RG invariance and a high-fidelity description of experimental data across diverse nuclear systems. 
It is widely held that cutoff dependence reflects missing higher-order interactions or complex many-body forces, suggesting that robust RG invariance may remain out of reach at the current level of EFT truncation~\cite{machleidt2023fbs}. 
In this work, we challenge this paradigm through a systematic RG study spanning from light nuclei to nuclear matter. 
We utilize a minimal chiral force constrained solely by few-body data, and employ a globally consistent regulator defined in terms of \emph{absolute} single-particle momenta~\cite{lu2022prl, liu2025epja, wu2025prc, zhang2025prd, wang2025prc, shi2026plb}. 
Originally introduced for mitigating lattice artifacts in nuclear lattice simulations, this regularization scheme has shown remarkable success in few-body systems, yielding cutoff-independent reproduction for the $^4$He binding energy within $100$ keV \cite{shi2026plb}. 
we now extend its application to substantially larger systems up to $^{40}$Ca and nuclear matter. 
Our results reveal that the choice of regulator, often dismissed as a technical nuance, plays a decisive role in simultaneously enhancing RG invariance and improving agreement with experiment.

\begin{figure}[h]
    \centering
        \includegraphics[width=1\linewidth]{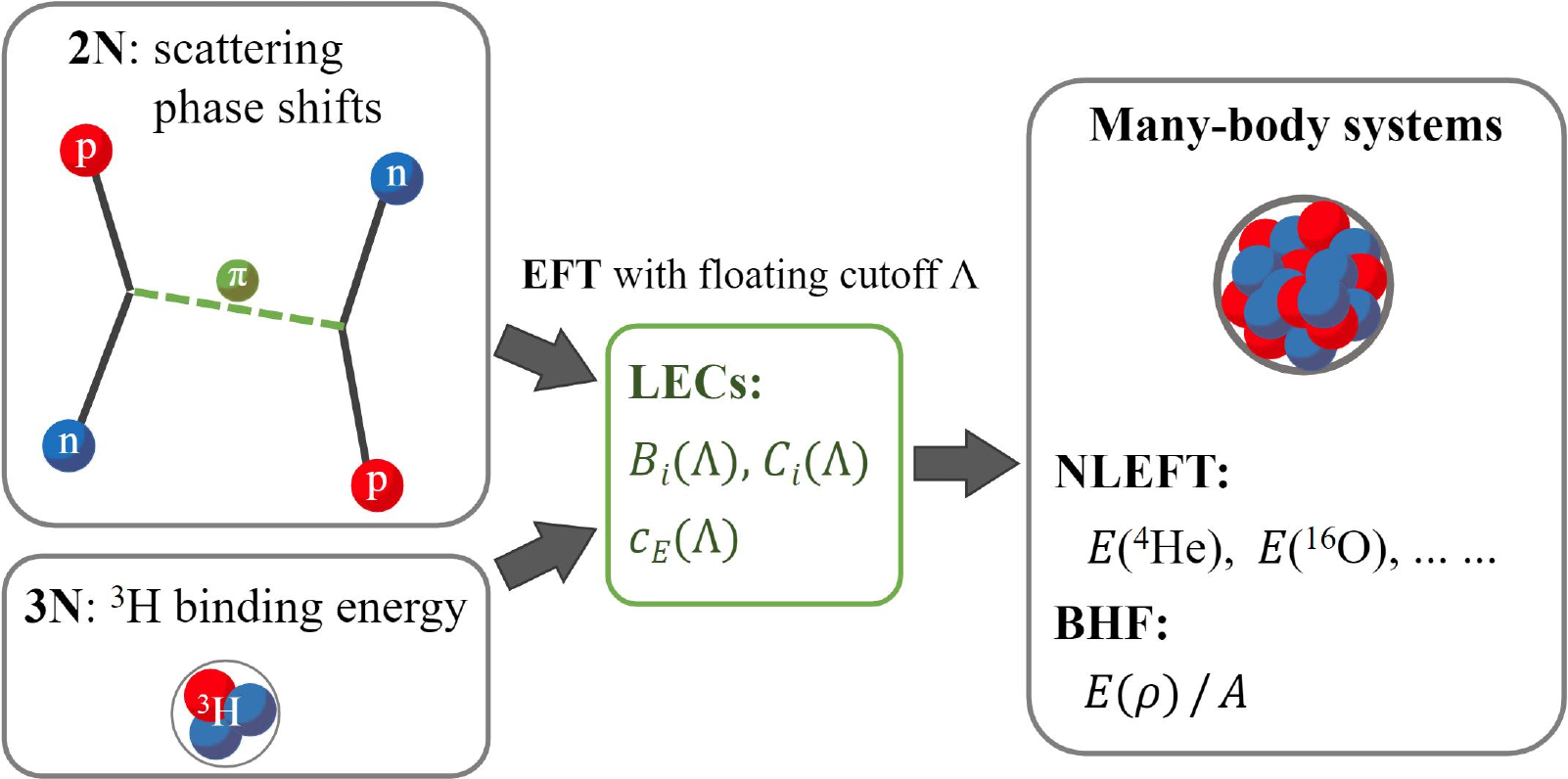}
        \caption{Schematic workflow. Contact LECs are calibrated using neutron-proton scattering data and the triton binding energy, then applied to many-body systems to predict binding energies. The calculation is repeated for different values of the cutoff $\Lambda$.}    
        \label{fig:workflow}
\end{figure} 
   
\textit{Method.}---The overall workflow of this study is sketched in Fig.~\ref{fig:workflow}. We construct a family of chiral interactions with variable momentum cutoff $\Lambda$. For each $\Lambda$, the low-energy constants (LECs) are determined by fitting to two- and three-nucleon data. The resulting interactions are then used as input for many-body calculations, yielding predictions for nuclear binding energies and the equation of state of symmetric nuclear matter (SNM). Finite nuclei are computed using Nuclear Lattice Effective Field Theory (NLEFT)~\cite{lee2004prc,lee2005prc,lee2009ppnp,lahde2019lnp}, while the Brueckner--Hartree--Fock (BHF) method is applied to uniform nuclear matter~\cite{day1967rmp,haftel1970npa,song1998prl}. 

    We employ the next-to-next-to-leading-order (N$^2$LO) chiral interaction introduced in Ref.~\cite{lu2022prl}, which consists of both two- and three-nucleon forces (2NFs and 3NFs):
\begin{align}
 V_{{\rm 2N}}= & \left[B_{1}+B_{2}(\bm{\sigma}_{1}\cdot\bm{\sigma}_{2})+C_{1}q^{2}+C_{2}q^{2}(\bm{\tau}_{1}\cdot\bm{\tau}_{2})\right.\nonumber \\
 & +C_{3}q^{2}(\bm{\sigma}_{1}\cdot\bm{\sigma}_{2})+C_{4}q^{2}(\bm{\sigma}_{1}\cdot\bm{\sigma}_{2})(\bm{\tau}_{1}\cdot\bm{\tau}_{2})\nonumber \\
 & +C_{5}\frac{i}{2}(\bm{q}\times\bm{k})\cdot(\bm{\sigma}_{1}+\bm{\sigma}_{2})+C_{6}(\bm{\sigma}_{1}\cdot\bm{q})(\bm{\sigma}_{2}\cdot\bm{q})\nonumber \\
 & \left.+C_{7}(\bm{\sigma}_{1}\cdot\bm{q})(\bm{\sigma}_{2}\cdot\bm{q})(\bm{\tau}_{1}\cdot\bm{\tau}_{2})\right]f_{{\rm abs}}(p_1^\prime, p_1, p_2^\prime, p_2)\nonumber \\
 & -\frac{g_{A}^{2}f_{\pi}(q^{2})}{4F_{\pi}^{2}}\left[\frac{(\bm{\sigma}_{1}\cdot\bm{q})(\bm{\sigma}_{2}\cdot\bm{q})}{q^{2}+M_{\pi}^{2}}+C_{\pi}^{\prime}\bm{\sigma}_{1}\cdot\bm{\sigma}_{2}\right](\bm{\tau}_{1}\cdot\bm{\tau}_{2}),  \nonumber \\
 V_{3\mathrm{N}} = & -\frac{3c_E}{F_\pi^4 \Lambda_\chi} 
 f_\mathrm{abs}(p_1^\prime, p_1, p_2^\prime, p_2, p_3^\prime, p_3),
\label{eq:V2N}
\end{align}
   where $\bm{p}_{1,2,3}$ and  $\bm{p}^\prime_{1,2,3}$ are incoming and outgoing
    momenta of individual nucleons, 
    $\bm{q} = \bm{p}^\prime-\bm{p}$ and $\bm{k} = (\bm{p}^\prime+\bm{p})/2$ 
    are momentum transfers with $\bm{p}^\prime = (\bm{p}_1^\prime - \bm{p}_2^\prime)/2$ and $\bm{p} = (\bm{p}_1 - \bm{p}_2)/2$ 
     the relative momenta, 
    $\bm{\sigma}_{1,2}$ and $\bm{\tau}_{1,2}$ are Pauli matrices 
    for spin and isospin, respectively.  
For the one-pion exchange potential (OPEP), $g_A$, $F_\pi$ and $M_\pi$  are the axial-vector coupling constant, pion decay constant and pion mass, respectively, with $C_\pi^\prime$  a constant counter term and $f_\pi(q^2)$ a local regulator for pion momentum, both of which are introduced to remove the singularities in the OPEP~\cite{reinert2018epja}.  
In the 3NF, $\Lambda_\chi$ is the chiral symmetry breaking scale~\cite{epelbaum2010epja} and $c_E$ is an adjustable dimensionless parameter.  
All contact terms are regulated by truncating \emph{absolute} single-particle momenta,
\begin{equation}\label{eq:fSP}
	f_\mathrm{abs}(p_1^\prime,p_1,\dots,p_A^\prime,p_A) = \prod_{i=1}^A
    \exp\left(-\frac{p^{\prime 2n}_i + p^{2n}_i}{2\Lambda^{2n}}
    \right), 
\end{equation}
with $\Lambda$ the momentum cutoff.
Here we take $n = 3$.
For finite nuclei, a perturbative static Coulomb force is included.  
Further technical details of the many-body methods and interactions  are provided in the Supplemental Material~\cite{supp}.

In this work, we keep the OPEP fixed and concentrate on the renormalization of the short-range contact terms. 
The LECs $B_{1-2}$ and $C_{1-7}$ are determined by fitting to empirical nucleon-nucleon (NN) scattering phase shifts up to a relative momentum of $p_{\mathrm{rel}} \leq 200~\mathrm{MeV}$~\cite{lu2016plb}, while the three-body LEC $c_E$ is adjusted to reproduce the experimental triton binding energy $E(^3\mathrm{H}) = -8.482~\mathrm{MeV}$.
In this momentum range, where $M_\pi \simeq p_{\mathrm{rel}} \ll 2M_\pi$, the OPEP in Eq.~(\ref{eq:V2N}) provides the dominant pion-pole contributions for NN scattering amplitudes. 
The two-pion exchange potentials, being analytic in this region, can be effectively absorbed into the contact terms and are therefore neglected~\cite{li2018prc,wu2025prc,liu2025epja}.

In previous EFT studies, two-body contact terms are conventionally regulated by truncating either relative momenta $\bm{p}$ and $\bm{p}^\prime$ for the convenience of partial-wave analysis~\cite{entem2003prc,epelbaum2005npa}, or the momentum transfer $\bm{q}$ to simplify many-body computations~\cite{gezerlis2013prl, lynn2017prc}, and many-body forces are often regulated using Jacobi momenta.
These regulators automatically preserve Galilean invariance.
Here, we compare the present \emph{absolute}-momentum regulator $f_{\rm abs}$ defined in Eq.~(\ref{eq:fSP}) with the widely adopted nonlocal \emph{relative}-momentum regulator $f_{\rm rel}$.
For the 2NFs, we have
\begin{align} 
f_\mathrm{abs} & = \exp\!\left(-\frac{p_1^{\prime  2n}+p_1^{2n}+p_2^{\prime 2n}+p_2^{2n}}{2\Lambda^{2n}}\right), \label{fSP2} \\ \nonumber
f_\mathrm{rel} & = \exp\!\left[-\frac{(\bm{p}_1^\prime-\bm{p}_2^\prime)^{2n}+(\bm{p}_1-\bm{p}_2)^{2n}}{(2\Lambda)^{2n}}\right].
\end{align}

In the two-nucleon (2N) center-of-mass frame, where $\bm{p}_1 = -\bm{p}_2$ and $\bm{p}_1^\prime = -\bm{p}_2^\prime$, the two regulators coincide, ensuring that fitting 2N forces yields identical LECs for both schemes. 
However, their behaviors diverge in many-body environments. Specifically, $f_{\mathrm{abs}}$ forbids scattering whenever any participating nucleon momentum exceeds $\Lambda$, whereas $f_{\mathrm{rel}}$ permits interactions as long as the relative momentum remains small. 
For nucleons scattering in a medium, $f_{\mathrm{abs}}$ thus imposes a more stringent restriction on the available intermediate phase space, leading to reduced scattering amplitudes and effectively weaker $G$-matrix elements in BHF calculations. 
This suppression systematically lowers the binding energies of many-nucleon systems relative to $f_{\mathrm{rel}}$.
Furthermore, recent RG studies for $A \leq 4$ systems demonstrate that the breaking of Galilean invariance by $f_{\mathrm{abs}}$ induces only negligible energy corrections that decay as inverse powers of $\Lambda$~\cite{shi2026plb}. 
Consequently, we neglect such regulator artifacts in this work.
Detailed discussion on Galilean invariance is provided in the Supplemental Material~\cite{supp}.

    \begin{figure} 
        \centering  
        \includegraphics[width=1\linewidth]{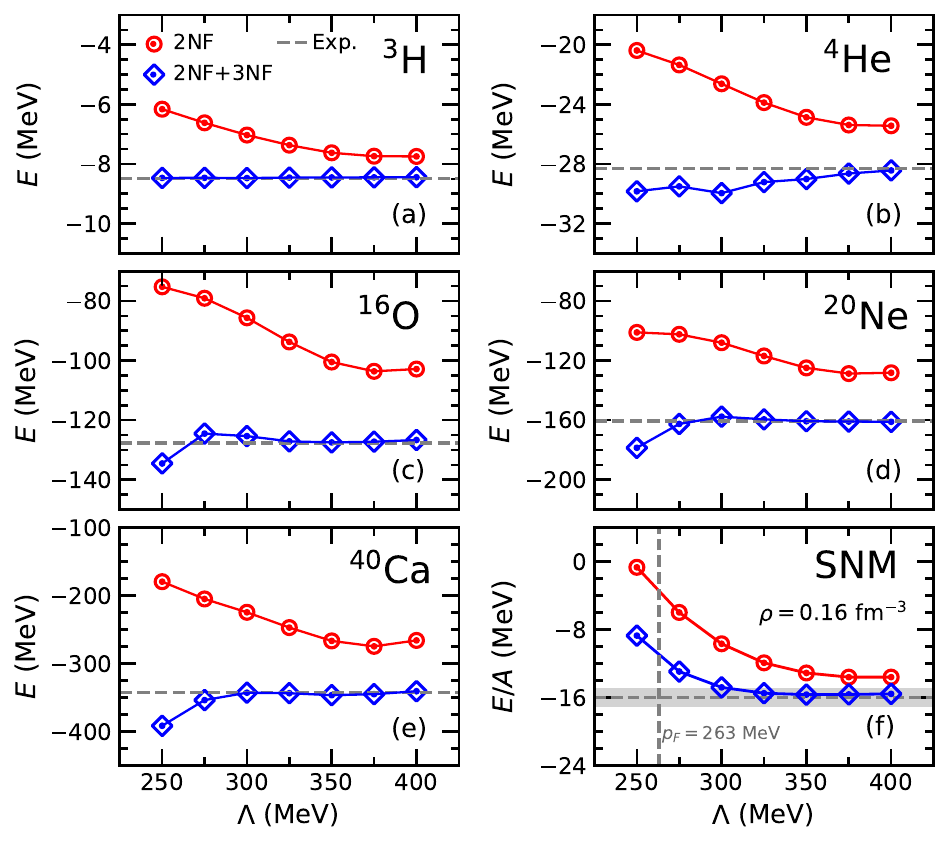} 
         \caption{(a-e) Binding energies of selected nuclei as functions of the cutoff $\Lambda$.
         Circles and diamonds denote results without and with 3NF, respectively.
         Dashed lines mark the experimental values.
         (f) Binding energy per nucleon for symmetric nuclear matter at empirical saturation density $\rho_0 = 0.16$~fm$^{-3}$.
         Vertical line represents the corresponding Fermi momentum.
         Numerical results are provided in Supplemental Materials~\cite{supp}.
            }\label{fig:LamRun}
    \end{figure}
    
\textit{Finite nuclei.}---To investigate regulator dependence, the momentum cutoff $\Lambda$ is tuned from $250$ to $400$ MeV. The calculations utilize a perturbative quantum Monte Carlo (ptQMC) algorithm that expands around a sign-problem-free Hamiltonian to second order, providing the precision necessary to resolve subtle regulator effects~\cite{lu2022prl}.
We set the lattice spacing to $a = 0.987$ fm, which corresponds to a sharp momentum cutoff $\Lambda_a = \pi/a \approx 628$ MeV. Since all considered values of $\Lambda$ remain well below $\Lambda_a$, lattice artifacts are effectively suppressed by the soft regulator $f_{\rm abs}$ and can be neglected. Consequently, the NLEFT results are expected to be consistent with those obtained from continuum-space \textit{ab initio} methods.

Fig.~\ref{fig:LamRun}(a--e) displays the binding energies of nuclei ranging from $^{3}$H to $^{40}$Ca as functions of the momentum cutoff $\Lambda$.
To elucidate the role of 3NFs, we compare the full results with calculations employing 2NFs alone.
The 2NF-only results systematically underbind these nuclei and exhibit a pronounced dependence on $\Lambda$, varying by about $20\%$--$30\%$ in the considered $\Lambda$-interval.
On the other hand, for the full 2NF+3NF interaction, the triton binding energy is reproduced by construction through our fitting procedure. Strikingly, the predictions for heavier nuclei [Fig.~\ref{fig:LamRun}(b--e)] demonstrate excellent RG invariance.
Particularly, for $\Lambda \geq 300~\mathrm{MeV}$, the binding energies remain essentially stable, with variations of only a few MeV, while simultaneously aligning with experimental data. Specifically, the $^{40}$Ca binding energy is predicted to be $-344.1(23)(22)$~MeV, where the first uncertainty accounts for the $\Lambda$ variation between $300$ and $400$~MeV and the second represents the many-body statistical error from NLEFT. 
This result is in excellent agreement with the experimental value of $-342.0$~MeV. 
Notably, the cutoff dependence of all nuclei observed in the 2NF-only sector is consistently absorbed by incorporating a single 3NF counterterm. 
Given that all LECs are fixed solely by $A \leq 3$ data, leaving no room for further readjustment, the effectiveness of this contact 3NF for heavier systems is particularly remarkable.

For $\Lambda = 400$ MeV, the 3NF typically contributes about $20\%$ to the total binding energy, representing $10\%$ of the total potential energy, similar in magnitude to chiral EFT results obtained with conventional regulators and consistent with the naive power couting estimation. Lowering the cutoff induces further three-body contributions to compensate for the variation, maintaining binding energy invariance as expected within the EFT picture.



\begin{figure}
    \centering
    \includegraphics[width=1\linewidth]{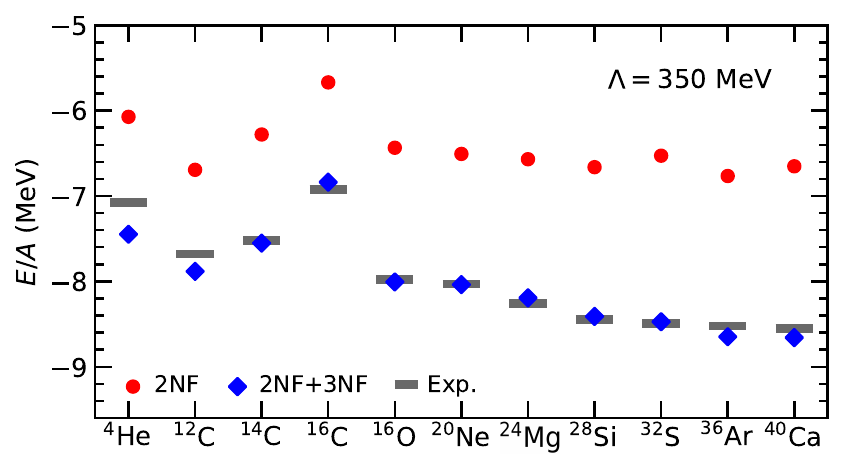}
    \caption{Binding energy per nucleon of selected nuclei calculated with $\Lambda=350$~MeV.
    Circles and diamonds denote results without and with 3NF, respectively.
    Grey bars indicate experimental values.}
        \label{fig:nucls350}
    \end{figure}

To further evaluate the predictive power of our framework, we extend the calculations to a broader range of nuclei. 
Fig.~\ref{fig:nucls350} displays the binding energies per nucleon for several even-even nuclei computed at a fixed cutoff $\Lambda = 350~\mathrm{MeV}$.
The 2NF-only results systematically underbind these nuclei by approximately $1$--$2~\mathrm{MeV}$ per nucleon, whereas the inclusion of 3NFs significantly improves the agreement with experimental data.
While the full interaction slightly overbinds $^{4}\mathrm{He}$ and $^{12}\mathrm{C}$, heavier nuclei are reproduced with high precision, with deviations typically at the 1\% level. 
The experimental trend across the carbon isotopes is also well captured. 
Given the established RG invariance, calculations at other cutoff values are expected to yield similar results. 
We further note that the residual overbinding in $^4\mathrm{He}$ is consistent with our previous studies and can be eliminated by restoring Galilean invariance~\cite{shi2026plb}.



\textit{Nuclear matter.}---The saturation properties of SNM have long served as a critical benchmark for nuclear interactions. 
Historically in BHF calculations, saturation points from 
phase-shift-equivalent two-body interactions 
typically lie on the so-called Coester line and fail to reproduce the empirical value~\cite{coester1970prc,li2006prc}. 
This longstanding issue remains within chiral EFT as well.
Previous \textit{ab initio} studies using BHF or many-body 
perturbation theory (MBPT) with soft 2NFs typically lead to severe 
overbinding and a lack of realistic  
saturation~\cite{hebeler2011prc,li2012prc,hu2017prc,drischler2019prl}. Under traditional regulation schemes imposed on 
Jacobi momenta, achieving a realistic description of SNM 
requires a delicate balance between the attraction provided by 
soft 2NFs and the repulsion from 3NFs by fine tuning LECs, 
especially $c_E$ and $c_D$~\cite{hebeler2011prc,drischler2019prl}.  

Fig.~\ref{fig:LamRun}(f) presents the energy per nucleon of SNM at the empirical saturation density $\rho_0 = 0.16\;\text{fm}^{-3}$ as a function of the cutoff $\Lambda$.
Clear evidence of RG-invariance emerges for the full 2N+3N interaction at $\Lambda \geq 300\;\text{MeV}$, where the converged energy aligns with the empirical value $E/A \approx -16$~MeV.
This agreement deteriorates for $\Lambda \leq 275\;\text{MeV}$, which can be attributed to the fact that the Fermi momentum $p_F \approx 263\;\text{MeV}$ at this density begins to exceed the momentum cutoff $\Lambda$, signaling the breakdown of the low-momentum expansion.

\begin{figure}[h]
        \centering
        \includegraphics[width=.8\linewidth]{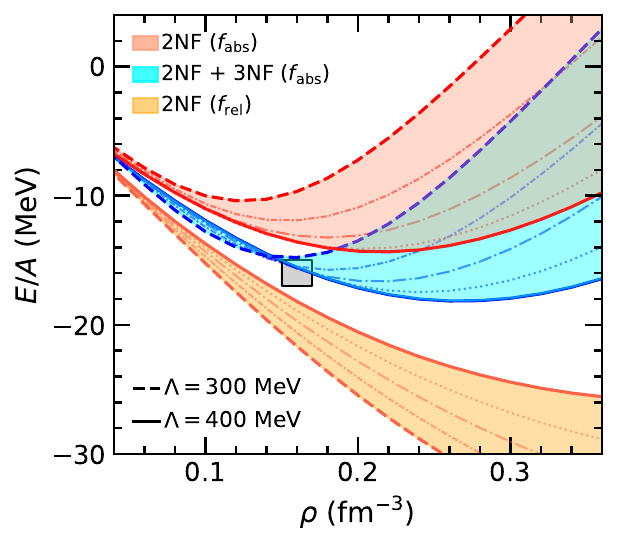}
        \caption{Binding energy per nucleon of symmetric nuclear matter.
        Red (Top) and blue (middle) lines denote results without and with 3NF, respectively.
        Orange (Bottom) lines indicate results calculated using 2NF with the same LECs and conventional \emph{relative}-momentum regulator.
        Each uncertainty bands are generated by varying cutoff $\Lambda$ from $300$~MeV (dashed lines) to $400$~MeV (solid lines).
        Grey patch shows the empirical saturation point.
        }\label{fig:snm_eos}
    \end{figure}

Fig.~\ref{fig:snm_eos} illustrates the equations of state (EOS) obtained within the range $300\;\text{MeV} \leq \Lambda \leq 400\;\text{MeV}$. We compare results using 2NFs alone with those incorporating the full 2N+3N interaction, where the variation in $\Lambda$ defines the corresponding uncertainty bands. 
With 2NFs only, the predicted $E/A$ forms a divergent band that widens significantly as density increases. 
Remarkably, the inclusion of 3NFs collapses this variation into a narrow, nearly $\Lambda$-independent band below the empirical saturation density $\rho_0$, providing robust evidence of RG-invariance in this regime. 
Above $\rho_0$, however, the uncertainty band of the 2N+3N results begins to expand. 
This behavior is anticipated within the EFT framework.
Since the 2NFs are calibrated using phase shifts only up to $p_{\rm rel} \leq 200$~MeV, the EFT description becomes inherently less reliable when the characteristic Fermi momentum $p_F$ significantly exceeds this scale. 
Consequently, we expect that incorporating higher-order interactions and constraining LECs with high-momentum data will substantially improve the description of high-density nuclear matter.

To further elucidate the role of the regulator, we directly compare the 2NF-only EOS's from $f_{\rm abs}$ and $f_{\rm rel}$ with the same LECs in Fig.~\ref{fig:snm_eos}. 
We find that $f_{\mathrm{rel}}$ leads to substantial overbinding and fails to yield a realistic saturation point across the investigated range of cutoffs, consistent with previous BHF studies utilizing \emph{relative}-momentum regularization in chiral EFT~\cite{li2006prc,li2012prc,hu2017prc}. 
In contrast, the \emph{absolute}-momentum regulator $f_{\mathrm{abs}}$ produces underbound EOS that exhibit well-defined saturation points even at the 2NF level. 
Notably, the overbinding of SNM is typically correlated with the systematic overbinding observed in medium-mass and heavy nuclei. While conventional approaches often rely on specific 3NF combinations to mitigate such overbinding, our results suggest an alternative framework that reduces the reliance on large 3NF corrections.
Specifically, the underbinding of the EOS at the 2NF level naturally leaves room for the attractive contribution of three-nucleon forces, aligning with the underbound triton result from 2NF alone shown in Fig.~\ref{fig:LamRun}(a).


    
    
 
\textit{Summary and perspective.}---
The regularization scheme plays a pivotal role in the renormalization group (RG) evolution of many-body nuclear systems. 
By employing an \emph{absolute} single-nucleon momentum regulator, we dramatically simplify the many-body calculation, achieving robust RG invariance and high-fidelity descriptions of binding energies up to the medium-mass region, utilizing only a single adjustable three-nucleon LEC. Our findings suggest that the long-standing ``impossible triangle'' of RG invariance, experimental precision, and structural simplicity is not an inherent limitation of EFT, but can be reconciled through a judicious  selection of the regulator. 
This perspective implies that the perceived necessity for increasingly complex higher-order interactions in many-body systems may, in part, be an artifact of conventional regularization choices.

While the \emph{absolute}-momentum regulator formally breaks Galilean invariance, our results show that this symmetry breaking is remarkably suppressed in many-body systems. The systematic restoration of this symmetry via Symanzik’s program~\cite{symanzik1983npb1,symanzik1983npb2} as demonstrated for $^4$He~\cite{shi2026plb} and $^{16}$O~\cite{supp} provides a clear path forward for further refinement~\cite{klein2018epja2, li2019prc, elhatisari2024nat}. 
The physical mechanism underlying this suppression remains a compelling open question that warrants deeper theoretical investigation.
The success of this framework across both NLEFT and BHF methodologies underscores its versatility. 
Our approach offers a practical and robust strategy for achieving RG invariance in EFT applications across nuclear and hadronic physics, quantum chemistry, and the study of cold atomic gases.

\textit{Acknowledgements.}---We thank members of the Nuclear Lattice 
Effective Field Theory Collaboration for insightful discussions.
B. L. thanks Dean Lee for critical reading of the manuscript.
C. W. thanks the enlightening discussion with Prof. X. L. Shang.
This work has been supported by NSAF No. U2330401 and National Natural Science 
Foundation of China with Grant Nos.~12547105,~12275259,~12375073~and~12405143.

\textit{Data availability.}---—The data are available from the authors upon 
request.

 
\providecommand{\noopsort}[1]{}\providecommand{\singleletter}[1]{#1}%

\clearpage

\appendix 
\setcounter{equation}{0}
\renewcommand{\theequation}{S\arabic{equation}} 

\setcounter{figure}{0}
\renewcommand{\thefigure}{S\arabic{figure}}
 
\setcounter{table}{0}
\renewcommand{\thetable}{S\arabic{table}}

\setcounter{page}{1}
\renewcommand{\thepage}{S\arabic{page}}  
\part{Supplemental Material}

\allowdisplaybreaks

In this appendix we present the methods and numerical details of the calculations.
Some of these materials have been published elsewhere as cited in the main text, here we include them for completeness.
We also provide a brief discussion about the Galilean invariance.

\section*{Low energy constants}
    We use the averaged mass $M_{\rm N} = 938.92$~MeV for both 
    proton and neutron.
    The two-nucleon force (2NF) adopted in this work consists of 
    long-range one-pion-exchange potential 
    (OPEP, $V_{1\pi}$) and zero-range contact terms 
    $V_{\rm ct} = V_{Q^0} + V_{Q^2}$.
    The OPEP reads 
    \begin{equation}\label{OBEP}
        V_{1\pi} = -\bm{\tau}_{1}\cdot\bm{\tau}_{2}
        \frac{g_A^2 f_{\pi}(q^2)}{4F_\pi^2}\left[
        \frac{(\bm{\sigma}_1\cdot\bm{q})(\bm{\sigma}_2\cdot\bm{q})}
        {q^2+M_\pi^2}+C_\pi'\bm{\sigma}_1\cdot\bm{\sigma}_2\right].
    \end{equation}
    Here the axial-vector coupling constant, decay constant and 
    pion mass are $g_A=1.287$, $F_\pi = 92.2$~MeV and 
    $M_\pi = 134.980$~MeV. 
    OPEP is regulated using semilocal scheme~\cite{sreinert2018epja},
    in which
    $f_\pi(q^{2})=\exp[-(q^{2}+M_{\pi}^{2})/\Lambda_\pi^{2}]$, 
    $\Lambda_\pi=300$~MeV and the counter term coefficient
    \begin{align*}
        C_\pi' =-\frac{1}{3}\Biggl[1-\frac{2M_\pi^2}{\Lambda_\pi^2}+
        2\sqrt{\pi}\frac{M_\pi^3}{\Lambda_\pi^3}\exp\left(
        \frac{M_\pi^2}{\Lambda_\pi^2}\right){\rm erfc}
        \left(\frac{M_\pi}{\Lambda_\pi}\right)\Biggr].
    \end{align*}    
    
    \begin{figure*}
    \centering
        \includegraphics[width=1\linewidth]{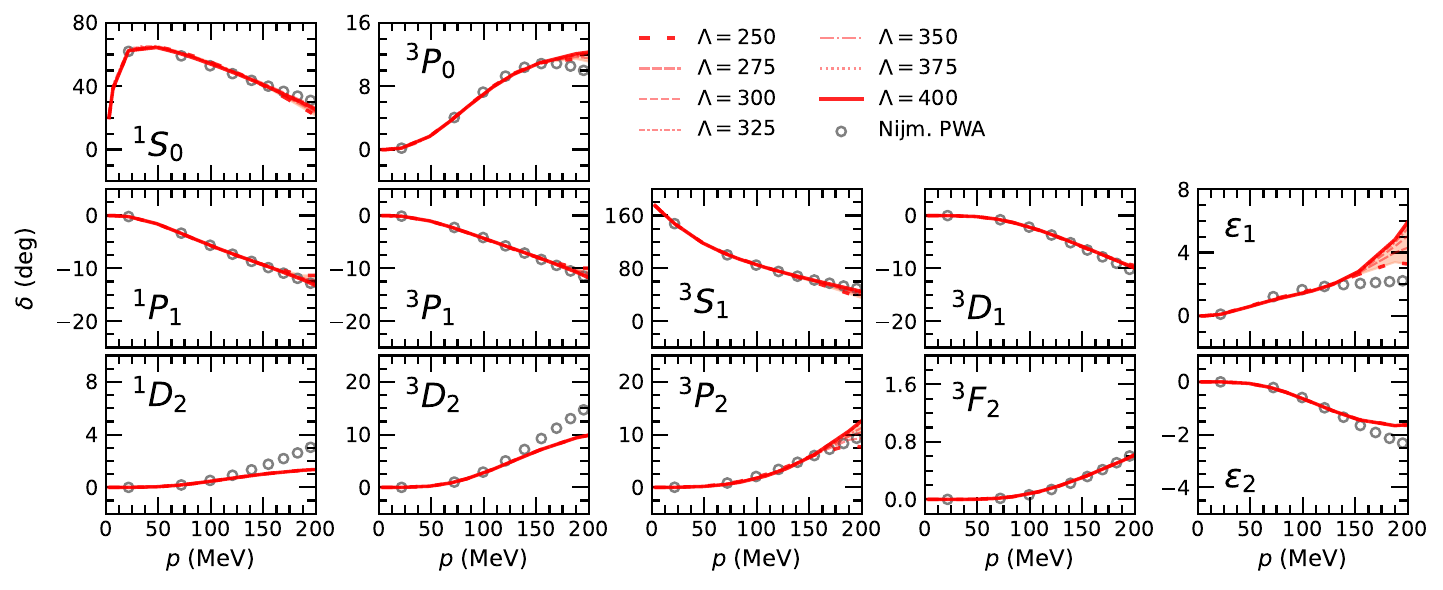}
        \caption{Nucleon-nucleon scattering phase shifts calculated 
        using the lattice chiral force with $\Lambda=250,~275,~\cdots,~400$ MeV. 
        The Nijmegen partial wave phase shifts (empty circles) are 
        also plotted for comparison. We add vertical lines for
        the upper limit of relative momentum $p=200$ MeV in the 
        fitting.}
        \label{fig:fishes}
    \end{figure*} 
    
    The contact terms of $V_{\rm ct}$ contains
    \begin{subequations}\label{Vct:LECs}
    \begin{align}
        V_{Q^0} &=  B_1+B_2\bm{\sigma}_{1}\cdot\bm{\sigma}_{2}
            \\  \nonumber
        V_{Q^2} &= q^2\left[ C_1+C_2 \bm{\tau}_1\cdot\bm{\tau}_2+
        C_3 \bm{\sigma}_1\cdot\bm{\sigma}_2 \right.
            \\ \nonumber
        &+ \left. C_4 (\bm{\sigma}_1\cdot\bm{\sigma}_2)
        (\bm{\tau}_1\cdot\bm{\tau}_2)\right]
        +C_5 i(\bm{q}\times\bm{k})\cdot(\bm{\sigma}_1+\bm{\sigma}_2)/2
            \\
        &  +(\bm{\sigma}_{1}\cdot\bm{q})
        (\bm{\sigma}_{2}\cdot\bm{q})
        (C_6 + C_7 \bm{\tau}_1\cdot\bm{\tau}_2).    
    \end{align}
    \end{subequations} 
    To obtain the low-energy constants (LECs) including
    $B_1,~B_2$ and $C_{1\text{--}7}$,
    we first determine the spectroscopic LECs in each partial wave.
    There are linear combinations
	\begin{subequations}\label{LECs:PWD}
	\begin{align}
      B_{{}^1S_0}=&B_1 - 3B_2, 
        \\  
      B_{{}^3S_1}=&B_1 + B_2, 
            \\  
		 C_{{}^1S_0}=&C_1+C_2-3C_3-3C_4-C_6-C_7, 
		 	\\  \nonumber  
		 C_{{}^3P_0}=&-2(C_1+C_2+C_3+C_4-C_5)/3 
            \\
           & -2(C_6+C_7),  
		    \\  \nonumber  
		 C_{{}^1P_1}=&-2(C_1-3C_2-3C_3+9C_4-C_6)/3 & 
            \\ 
            & -2C_7, 
		 	\\  
		 C_{{}^3P_1}=&-2(C_1+C_2+C_3+C_4)/3+C_5/3 
            \\ 
            & -4(C_6+2C_7)/3, 
		 	\\
		 C_{{}^3S_1}=&C_1-3C_2+C_3-3C_4+C_6/3-C_7, 
                \\
		 C_{SD}=&-2\sqrt{2}C_6/3+ 2\sqrt{2}C_7,  
		 	\\
	   C_{{}^3P_2}=&-2(C_1 + C_2 + C_3 + C_4)/3-C_5/3.
	\end{align}		
	\end{subequations} 
    Corresponding partial-wave-decomposed contact terms 
    are given by
    \begin{subequations}\label{Vct:Q2}
    \begin{align}
        &\langle S|V_{Q^0}|S\rangle=B_S 
        f_{\rm abs}(p'_1,p_1,p_2',p_2),
            \\
    	& \langle S|V_{Q^2}|S\rangle=C_S~(p'^2 + p^2) 
        f_{\rm abs}(p'_1,p_1,p_2',p_2), 
             \\
        &\langle S|V_{Q^2}|D\rangle= C_{SD}~p^2 
        f_{\rm abs}(p'_1,p_1,p_2',p_2), 
            \\
    	&\langle P|V_{Q^2}|P\rangle = C_P~p'p 
        f_{\rm abs}(p'_1,p_1,p_2',p_2).
    \end{align} 
    \end{subequations} 
    
    We compute the scattering phase shifts $\delta$ at given 
    relative momentum $p$ using lattice spherical wall method 
    with angular momentum projection and auxiliary 
    potential~\cite{slu2016plb}. 
    The spectroscopic LECs are determined by optimizing  
    \begin{equation}
        \chi^2 = \frac{1}{N_p}\sum_p^{N_p}\left[
        \frac{\delta(p)-\delta_\mathrm{Nijm}(p)}
        {\Delta \delta_\mathrm{Nijm}(p)}\right]^2,
    \end{equation} 
    with $\delta_\mathrm{Nijm}$ and $\Delta\delta_\mathrm{Nijm}$ 
    being the Nijmegen partial-wave phase shifts and corresponding
    uncertainties~\cite{sstoks1994prc} up to a relative momentum 
    $p=200$~MeV (corresponding to laboratory energy 
    $E_{\rm lab} = 85$ MeV). 
    Then the $B_i$, $C_i$ can be obtained by solving the  
    linear Eqs.~\eqref{LECs:PWD}. 
 
    By construction, the fitted LECs reproduce  
    the $S$- and $P$-wave phase shifts within the calibration 
    range $p \leq 200$ MeV for cutoffs $\Lambda \in [250, 400]$ MeV.
    For higher partial waves ($D, F$, etc.), where no contact terms 
    appear at this chiral order, the phase shifts are determined 
    solely by the OPEP without cutoff dependence.
    While the axial-vector coupling $g_A$ could theoretically be 
    treated as a running parameter, our sensitivity analysis 
    indicates that the physical value $g_A = 1.287$ consistently 
    provides the optimal fit across the considered cutoff range. 
    This suggests that the long-range OPEP contribution remains 
    robust and can be well separated against the contact terms  
    under the chosen regularization scheme.
    
\begin{figure*}
    \centering
        \includegraphics[width=1\linewidth]{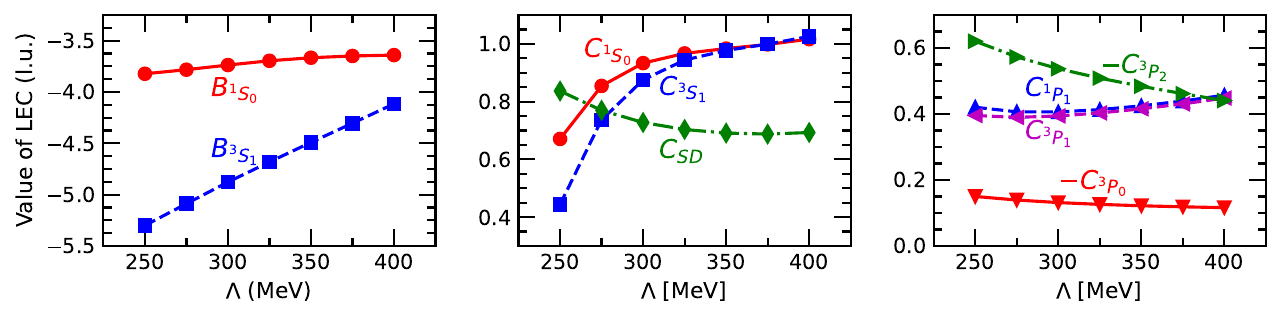}
        \caption{The LECs of $V_{\rm ct}$ in lattice unit running 
        with cutoffs (the physical value of a LEC is obtained by 
        multiplying proper powers of lattice spacing $a$ ).}
        \label{fig:lecrunning}
    \end{figure*}

    The fitted LECs are depicted in 
    Fig.~\ref{fig:lecrunning}.  
    As the cutoff $\Lambda$ is varied from $250$ to $400$~MeV, each 
    spectroscopic LEC evolves continuously to compensate for the 
    truncated high-momentum physics. This smooth ``running'' is a 
    hallmark of a consistent Renormalization Group (RG) evolution. 
    
    Compared with other high-precision chiral interactions,
    our construction lacks the two-pion exchange potentials 
    (TPEPs) both in the two-nucleon and three-nucleon sectors. 
    Since TPEPs contributions can be effectively absorbed into the 
    LECs in the contact terms of similar spin-momentum structure.
    It was long known that the contact three-nucleon forces 
    (3NFs) with undetermined LECs $c_D$ and $c_E$ at N$^2$LO can 
    not be uniquely fixed by $^3$H and $^4$He binding energies
    due to the Tjon line correlation~\cite{stjon1975plb}.
    Therefore we employ a simplified 3NF consisting solely of the fully 
    contact term governed by a single LEC, $c_E$. 
    For each cutoff $\Lambda$, after the LECs in 2NFs obtained,
    $c_E$ is uniquely determined by reproducing the experimental 
    triton binding energy, $E_\text{Exp.}(^3\text{H}) = -8.482$ MeV.
    
    On the discrete lattice, the triton ground state is obtained 
    via sparse matrix diagonalization using the Lanczos algorithm. 
    To remove periodic boundary effects, we perform an 
    infinite-volume extrapolation~\cite{smeissner2015prl}
    \begin{equation}\label{3Nextrap}
        E_{^3{\rm H}}(L) = E_{^3{\rm H}}(+\infty) + 
        \frac{\mathcal{C}}{(q_{\rm B} L)^\frac{3}{2}} \exp
        \left(-2q_{\rm B} L/\sqrt{3}\right),
    \end{equation}
    where $E_{^3{\rm H}}(+\infty)$ is the binding energy and
    $q_{\rm B} = \sqrt{-E_{^3{\rm H}}(+\infty) M_{\rm N}}$.
    \begin{figure*}
    \centering
        \includegraphics[width=.75\linewidth]{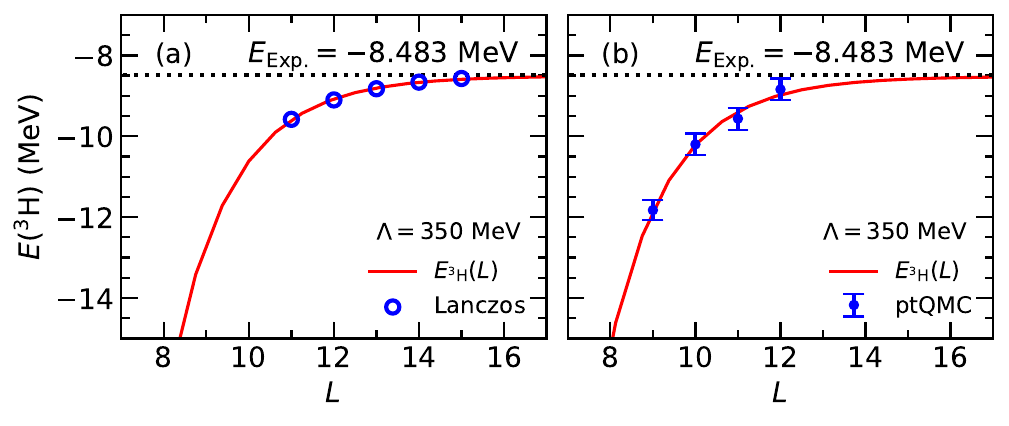}
        \caption{(a) Triton binding energy calculated by
        Lanczos method with infinite-volume extrapolation.
        (b) Triton binding energy calculated by
        perturbation Monte-Carlo method with 
        infinite-volume extrapolation.}
        \label{fig:finvol}
    \end{figure*} 
    In Fig.~\ref{fig:finvol}(a), we present infinite extrapolations for the triton binding energy. 
    We first using Lanczos method to solve the binding energy 
    exactly inside boxes of varying $L$. 
    We then perform the infinite-size extrapolation based on 
    Eq.~\eqref{3Nextrap} to get $E_{^3\text{H}}(+\infty)$.
    By construction, the extrapolated $E_{^3\text{H}}(+\infty)$ with three-body force included
    exactly reproduce the experimental value.
    For comparison, in Fig.~\ref{fig:finvol}(b) we also 
    present triton binding energy calculated by
    perturbation Monte Carlo (ptQMC) method~\cite{slu2022prl} 
    with the same extrapolation. 
    Although associated with statistical uncertainty, 
    the ptQMC results with extrapolation can also reproduce 
    triton binding energy. 
    In next section, ptQMC method will be employed to 
    explore heavier nuclei.

\section*{Perturbative quantum Monte Carlo method}
    The ground-state binding energies of nuclei from ${}^4$He 
    to $^{40}$Ca are computed using ptQMC method, 
    which starts from a zeroth order Hamiltonian respecting the 
    Wigner SU(4) symmetry~\cite{swigner1937pr}
    \begin{equation}\label{eq:SU(4)Hamiltonian}
        H_0=K+\frac{1}{2}C_{\rm SU4}\sum_{\bm{n}}
        \no{\tilde{\rho}^2(\bm{n})}.
    \end{equation}
    Here $\bm{n}=(n_{x},n_{y},n_{z})$ denotes the lattice coordinates, 
    $K$ represents the kinetic energy 
    \begin{equation}
        K = \sum_{\bm{n},i} a^\dagger (\bm{n},i)
        \frac{-\nabla^2}{2M_{\rm N}} a(\bm{n},i),
    \end{equation}
    with $a$ ($a^\dagger$) being the annihilation (creation) 
    operator, $i$ being the joint spin-isospin index,
    and the $::$ symbol indicating normal ordering. 
    The smeared density operator $\tilde{\rho}(\bm{n})$ is defined as
    \begin{align}\nonumber
        \tilde{\rho}(\bm{n})=&\sum_{i}\tilde{a}^\dagger(\bm{n},i)
        \tilde{a}(\bm{n},i) 
            \\
        &+s_{\rm L}\sum_{|\bm{n}'-\bm{n}|=1}
        \sum_{i}\tilde{a}^\dagger(\bm{n},i)\tilde{a}_(\bm{n}',i),
    \end{align}
    with $s_{\rm L}$ being the local smearing strength.
    The smeared annihilation operator is defined as
    \begin{equation}\label{NLsmearing}
        \tilde{a}(\bm{n},i)=\sum_{\bm{n}'}g(\bm{n}-\bm{n}')a(\bm{n}',i),
    \end{equation}
    with 
    $g(r)=(\Lambda_{\rm NL}/\sqrt{2\pi})^3
    \exp(-\Lambda_{\rm NL}^2r^2/2)$.

    Both $s_{\rm L}$ and $\Lambda_{\rm NL}$ control the range 
    of the interaction. A proper calibration of $s_{\rm L}$ and 
    $\Lambda_{\rm NL}$ together with $C_{\rm SU4}$ is capable to
    roughly reproduce the binding energies from light to 
    medium-mass nuclei. In our lattice calculation, we use the parameter set 
    $s_{\rm L}=0.182$,   $\Lambda_{\rm NL} = 300$~MeV.
    and $C_{{\rm SU4}}=-3.7912\times10^{-5}$ MeV$^{-2}$.
    
 	As a noteworthy feature, the absolute-momentum regulator
    $f_{\rm abs}$ is separable
    \begin{equation}
        f_{\rm abs}(p_1',p_1,~\cdots~p_A',p_A) = 
        f_\Lambda(p_1')f_\Lambda(p_1)\cdots
        f_\Lambda(p_A')f_\Lambda(p_A),
    \end{equation}
    with $f_\Lambda(p) = \exp[-p^2/(2\Lambda^6)]$.  
    Given the Fourier transform 
    $\tilde{f}_\Lambda(\bm{r}) = {\rm FT}[f_\Lambda(p)]$,
    the regulated annihilation operator on lattice get ``dressed'' by 
    $\tilde{f}_\Lambda$ through \textit{convolution theorem}
 	\begin{align}\nonumber  
        a_\Lambda(\bm{n},i) =& {\rm FT} \left[f_\Lambda(p) 
        a(\bm{p},i)\right]
            \\
        = &\sum_{\bm{n}'} \tilde{f}_\Lambda(\bm{n}-\bm{n'}) 
        a(\bm{n}',i).
 	\end{align}	
    Therefore $\tilde{f}_\Lambda(\bm{n})$ 
    plays a similar role as nonlocal smearing $g(\bm{n})$ in 
    Eq.~\eqref{NLsmearing}.
    
    The ground state corresponding to $H_0$ can be obtained 
    by imaginary time projection 
    \begin{equation}\label{Psi0}
        |\Psi_0\rangle\propto\lim_{L_t\rightarrow +\infty}
        M^{L_t/2}_0 |\Phi_T \rangle,
    \end{equation}
    where $|\Phi_T\rangle$ denotes an initial trial ground state
    (a Slater determinant composed of harmonic oscillator wave functions), 
    $M_0 = \no{\exp\left(-a_t H_0\right)}$ is the transfer matrix 
    and $a_t^{-1} = 1000$ MeV is the temporal interval.
    Wigner-SU(4) Hamiltonian is sign-problem-free in the Monte
    Carlo simulation~\cite{slu2019plb} and corresponding 
    solutions are very accurate. However, the chiral Hamiltonian 
    \begin{equation}
 		  H = K + V_\mathrm{2N} + V_\mathrm{3N} +  
        V_\mathrm{Cou}
 	\end{equation}
    breaks the SU(4) symmetry and induces severe sign problem.
    Thus $H$ will be treated perturbatively.
    For lattice calculation, $V_{\rm 2N}$ are given in lattice 
    coordinate.
    \begin{align}\nonumber
        V_{\rm 2N} = &-\frac{g_A^2}{8F_\pi^2}\sum_{\bm{n},I}
        \sum_{S_1S_2} \rho_\Lambda(\bm{n},S_1,I)
        G_\pi(\bm{n}_1-\bm{n}_2,S_1,S_2)
            \\ \nonumber
        &\times \rho_{\Lambda}(\bm{n},S_2,I) + \frac{1}{2}
        \sum_{\bm{n}}:\!\!\left[B_1\rho^2_\Lambda(\bm{n}) +B_2
        \sum_S \rho^2_\Lambda(\bm{n},S) \right. 
            \\ \nonumber 
        &-C_1\rho_\Lambda(\bm{n})\Delta\rho_\Lambda(\bm{n})- 
        C_2\sum_I\rho_\Lambda(\bm{n},I)\Delta\rho_\Lambda(\bm{n},I)
            \\ \nonumber 
        &-C_3\sum_S\rho_\Lambda(\bm{n},S)\Delta\rho_\Lambda(\bm{n},S)
            \\ \nonumber 
        &-C_4\sum_{SI}\rho_\Lambda(\bm{n},S,I)\Delta\rho_\Lambda
        (\bm{n},S,I)
            \\ \nonumber 
	    &+\left. C_5\sum_S\rho_{\Lambda,{\rm SO}}(\bm{n},S)
        \rho_\Lambda(\bm{n},S)+C_6\rho_{\Lambda,{\rm T}}(\bm{n})
        \rho_{\Lambda,{\rm T}}(\bm{n})\right.
            \\ 
        & + \left. C_7\sum_I 
	    \rho_{\Lambda,{\rm T}}(\bm{n},I)
		\rho_{\Lambda,{\rm T}}(\bm{n},I)\right]\!\!:.
    \end{align}
    The pion propagator $G_\pi$ is defined as 
    \begin{align}
        G_\pi(\bm{n},S_1,S_2) = \sum_{\bm{q}}
        e^{i\bm{q}\cdot\bm{n}} f_\pi (q)
        \left(\frac{q_{S_1}q_{S_2}}{q^2 + M_\pi^2} + 
        C'_\pi\delta_{S_1S_2}\right).
    \end{align}
    Various densities are given by
    \begin{subequations}
	\begin{align}
        \rho_\Lambda(\bm{n},S,I)=&\sum_i a^\dagger_\Lambda(\bm{n},i)
        [\sigma_S\otimes \tau_I]_{ii'}a_\Lambda(\bm{n},i'),
            \\  \nonumber 
		\rho_{\Lambda,{\rm SO}}(\bm{n},S)=&\frac{i}{2}\sum_{ll'S}
		\epsilon_{Sll'}\sum_{ii'} \partial_l a^\dagger_\Lambda(\bm{n},i)
        [\sigma_S]_{ii'}    
            \\
        & \times \partial_{l'}a_\Lambda(\bm{n},i'),
		      \\  
		\rho_{\Lambda,{\rm T}}(\bm{n},I)=&\sum_{i'i}
		a_\Lambda^\dagger(\bm{n},i)\left[\bm{\sigma}\cdot
		(\bm{\nabla}-\overleftarrow{\bm{\nabla}})\otimes \tau_I
        \right]_{ii'}
            \\  \nonumber 
        & \times a_\Lambda(\bm{n},i').
	\end{align}
    \end{subequations}
    Here $l$, $S$, $I$ take the values of $1,~2,~3$ for 
    $x$, $y$, $z$ directions. We also introduce the Pauli
    matrices $\sigma_{S=0} = \tau_{I=0} = 1_{2\times 2}$ and 
    denote $\rho_\Lambda(\bm{n}) = \rho_\Lambda(\bm{n},0,0)$, 
    $\rho_\Lambda(\bm{n},S) = \rho_\Lambda(\bm{n},S,0)$ and 
    $\rho_\Lambda(\bm{n},I) = \rho_\Lambda(\bm{n},0,I)$.
    The  three-body contact interaction is given by
    \begin{equation}
    	V_{\rm 3N} = -\frac{ c_E}{2F_\pi^4 \Lambda_\chi} 
        \sum_{\bm{n}}\no{\rho^3_\Lambda(\bm{n})}.
    \end{equation}
    $V_{\rm Cou}$ denotes Coulomb potential, given by 
    \begin{equation}
        V_\mathrm{Cou} = \frac{1}{2} \sum_{\bm{n}_1 \bm{n}_2}
        \frac{\alpha}{\mathrm{max}(|\bm{n}_1-\bm{n}_2|,0.5)}
        \no{\rho_\mathrm{p}(\bm{n}_1)\rho_\mathrm{p}(\bm{n}_2)}.
    \end{equation}
    where $\alpha=1/137$ is the fine structure constant and 
    $\rho_{\rm p}(\bm{n})$ is the proton density operator.
    \begin{figure*}
    \centering
        \includegraphics[width=1\linewidth]{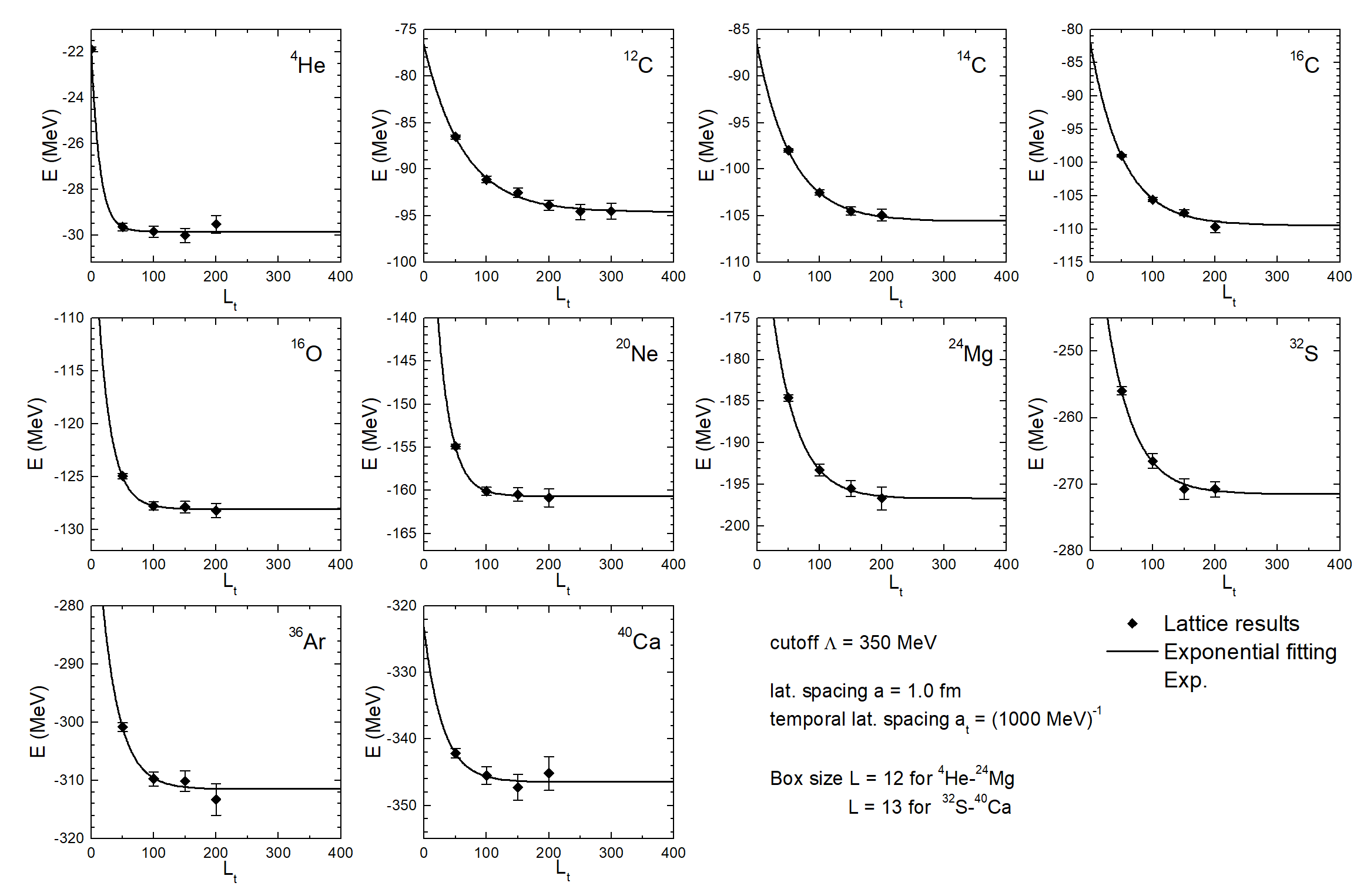}
        \caption{Infinite-time extrapolation for the ptQMC results.
        In each panel, the dots with error bars are results of 
        ptQMC calculations and the solid curve is the 
        extrapolation function of Eq.~\eqref{E(Lt)} after fitting. }  
        \label{fig:finLt}
    \end{figure*} 
    
    One can construct the perturbation with respect
    to the ground state of Eq.~\eqref{Psi0} as
 	\begin{align}\label{Psid1}
 		|\delta \Psi \rangle\propto &\lim_{L_t\rightarrow +\infty}
        \sum_{k=1}^{L_t/2}  M_0^{L_t/2-k}(M-M_0)M_0^{k-1} 
        |\Phi_T\rangle,
 	\end{align} 
    with $M = \no{\exp(-a_t H)}$.   
    One can also construct the Hamiltonian perturbation 
    $\delta H = H - H_0$, correspondingly, 
    the ground-state binding energy can be expanded 
    as $E = E_0 + \delta E_1 + \delta E_2$,  where 
 	\begin{subequations}
 	\begin{align}
 		E_0  =& \langle \Psi_0|H_0|\Psi_0 \rangle
            /\langle \Psi_0|\Psi_0 \rangle, 
 			\\
        \delta E_1  =& \langle \Psi_0|\delta H|\Psi_0 \rangle
            /\langle \Psi_0|\Psi_0 \rangle,
            \\
        \delta E_2  =&\mathrm{Re}\langle \Psi_0|\delta H - \delta E_1
        |\delta \Psi\rangle/\langle \Psi_0|\Psi_0 \rangle.
 	\end{align}  
 	\end{subequations}

    For saving the computational resources,  we have examined 
    the finite-volume effect with lattice chiral force of
    $\Lambda=350$~MeV and find the box size of $L=12,~13$ 
    are sufficiently close to the infinite limit. Another
    finite-size issue related to lattice calculations concerns 
    the finite value of imaginary time $L_t$~in Eqs.~\eqref{Psi0} 
    and~\eqref{Psid1}. The calculated energy inherently depends 
    on the finite value of $L_t$, and the extrapolation to 
    infinite is given by 
    \begin{equation}\label{E(Lt)}
		E(L_t)= E_0 + \mathcal{C} e^{-L_t\Delta E_1}
	\end{equation}
    where $E_0$ denotes the ground-state binding energy.
    In Fig.~\ref{fig:finLt}, we present the illustrations of 
    infinite-time extrapolation using the representative chiral
    force of $\Lambda=350$~MeV.

    In Tabs.~\ref{tab:EvsL} and~\ref{tab:E350}, the 
    results with uncertainties associated with both Monte Carlo 
    simulation and infinity-time extrapolation are given in detail. 
    Besides, rows 1--10 in Tab.~\ref{tab:EvsL} also give the 
    values of LECs in lattice unit.

The calculated binding energy of $^{40}\text{Ca}$ is presented in the main text with two distinct uncertainties. The central value is obtained by averaging the results across the five cutoff scales from $\Lambda = 300$ to $400$ MeV. The first uncertainty represents the standard deviation over this range, reflecting the cutoff dependence. The second uncertainty is the root-mean-square (RMS) of the combined Monte Carlo and extrapolation errors.

    \begin{table*}[t]
    \renewcommand{\arraystretch}{1.2}
    \setlength{\tabcolsep}{7.5pt} 
    \centering 
    \caption{LECs and binding energies of selected nuclei 
        versus the cutoff variation.
        Rows 1--10 give the LECs in the lattice chiral force 
        with $\Lambda=250\text{--}400$ MeV.
        For each row showing the binding energies of 
        $^3$H, $^{4}$He, $^{16}$O, $^{20}$Ne and $^{40}$Ca, 
        the first line gives the binding energies calculated 
        with $V_{\rm 2NF} + V_{\rm Cou}$, while the second line 
        shows the results obtained from 
        $V_{\rm 2NF} + V_{\rm 3NF} + V_{\rm Cou}$. 
        Numbers in the parentheses denote uncertainties from 
        Monte Carlo simulation and infinity-time extrapolation. 
        LECs are given in lattice unit and all energies are in MeV.}
    \begin{tabular}{c|rrrrrrrr}
    \toprule 
    $\Lambda$ (MeV)&$250$&$275$&$300$&$325$&$350$
                   &$375$&$400$&Exp.\\  
    \midrule  
	$B_1$ &-4.931&-4.762&-4.592&-4.435&-4.285&-4.140&-3.997\\
    $B_2$ &-0.369&-0.326&-0.285&-0.246&-0.206&-0.164&-0.119\\
    $C_1$ &0.363&0.432&0.454&0.456&0.449&0.440&0.432\\
    $C_2$ &0.063&0.011&-0.017&-0.032&-0.041&-0.048&-0.056\\
    $C_3$ &-0.002&-0.024&-0.033&-0.039&-0.042&-0.047&-0.051\\
    $C_4$ &0.008&-0.029&-0.049&-0.060&-0.067&-0.073&-0.078\\
    $C_5$ &0.997&0.939&0.901&0.875&0.856& 0.841&0.829\\
    $C_6$ & 0.024&0.015&0.007&0.002&-0.002&-0.004&-0.005\\
    $C_7$ &-0.288&-0.267&-0.255&-0.248&-0.245&-0.245&-0.247\\   
    $c_E$&5.170&2.763&1.538&0.890&0.561&0.412&0.380 &\\
        \midrule  
   \multirow{2}{*}{$E(^3$H)} 
    &-6.174&-6.634&-7.052&-7.392&-7.641
    &-7.771&-7.781&\multirow{2}{*}{-8.482} \\
    &-8.482&-8.482&-8.482&-8.482&-8.482&-8.482&-8.482& \\
    \midrule
    \multirow{2}{*}{$E(^4$He)} 
    & -20.4(3)&-21.4(3)&-22.6(3)&-23.9(3)&-24.9(3)
    &-25.4(3)&-25.4(3)&\multirow{2}{*}{-28.3} \\
    &-29.8(7)&-29.5(6)&-29.9(4)&-29.2(4)&-29.0(4)
    &-28.6(4)&-28.4(4)&\\ 
    \midrule
    \multirow{2}{*}{$E(^{16}$O)} 
    &-77.1(7)&-78.8(8)&-85.4(7)&-93.8(7)&-100.8(7)
    &-104.3(7)&-104.1(8)& \multirow{2}{*}{-127.6} \\
    &-135.8(20)&-124.8(14)&-124.5(11)&-126.3(9)&-127.3(8)
    &-128.1(8)&-128.1(8)& \\ 
    \midrule
    \multirow{2}{*}{$E(^{20}$Ne)}
    &-100.8(16)&-101.9(14)&-107.7(13)&-116.7(13)
    &-124.6(13)&-128.4(13)&-128.0(15)&\multirow{2}{*}{-160.6}\\
    &-178.7(30)&-162.5(21)&-157.8(16)&-159.6(14)
    &-160.9(12)&-160.8(12)&-161.1(15)& \\
    \midrule
    \multirow{2}{*}{$E(^{40}$Ca)} 
    &-180.3(24)&-204.5(19)&-224.6(17)&-247.6(17)&-266.8(17)
    &-274.3(17)&-266.1(18)&\multirow{2}{*}{-342.0}\\     
    &-391.6(62)&-355.8(40)&-346.5(28)&-341.6(24)&-346.4(15)
    &-341.3(20)&-344.7(20)& \\    
    \bottomrule	 
    \end{tabular}
    \label{tab:EvsL}
    \end{table*}

    \begin{table}[h]
    \renewcommand{\arraystretch}{1.2}
    \setlength{\tabcolsep}{10pt} 
    \centering 
    \caption{Binding energy of selected nuclei calculated by 
    the lattice chiral force with $\Lambda=350$~MeV. 
    Notations are same as Tab.~\ref{tab:EvsL}. }
    \begin{tabular}{c|rrr}
    \toprule
        &$E_\mathrm{2NF}$&$E_\mathrm{2NF+3NF}$&Exp.\\  
    \midrule  
        {$^4$He} &-24.9(4)&-29.0(4)&-28.3\\
        
        {$^{12}$C}&-80.4(7)&-94.6(4)&-92.2\\
        
        {$^{14}$C} &-87.9(4) &-105.6(3)&-105.3\\
        
        {$^{16}$C}&-90.7(14)&-109.4(12)&-110.8\\
        
        {$^{16}$O} &-100.8(7)&-127.3(8)&-127.6\\
        {$^{20}$Ne}&-124.6(13)&-160.9(12)&-160.6\\
        
        {$^{24}$Mg} &-157.7(11)&-196.7(4)&-198.3\\
        
        {$^{28}$Si}&-186.7(14)&-235.2(25)&-236.5\\
        
        {$^{32}$S}&-208.9(15)&-271.4(8)&-271.8\\
        
        {$^{36}$Ar} &-243.6(43)&-311.4(18)&-306.7\\
        
        {$^{40}$Ca}&-266.8(17)&-346.4(15)&-342.1\\
    \bottomrule	 
    \end{tabular} \label{tab:E350}
    \end{table}

\begin{figure*}[t]
    \centering
        \includegraphics[width=1\linewidth]{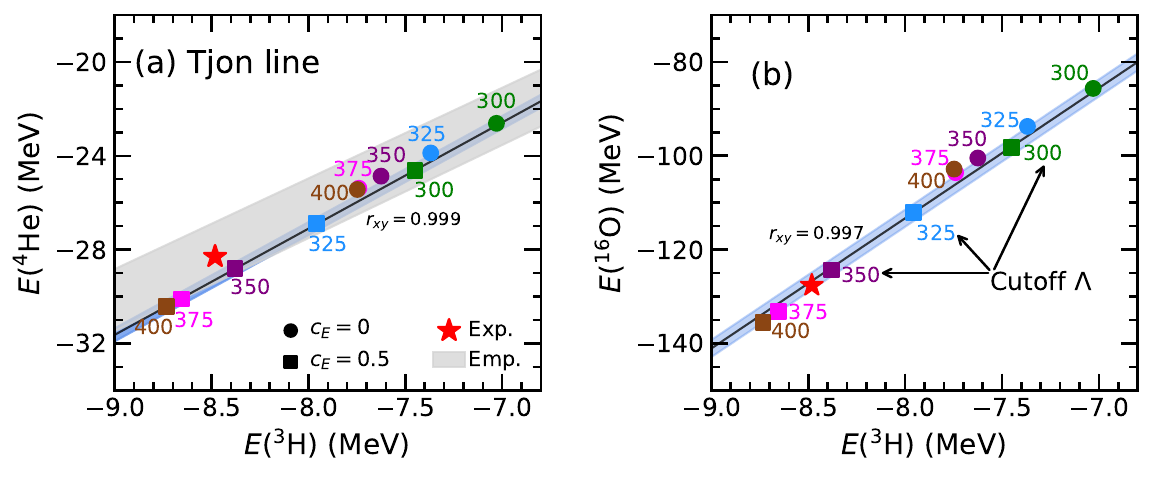}
        \caption{(a) The Tjon line (band) of $^3$H and $^4$He
        binding energies calculated by varying cutoffs 
        from 300~MeV to 400~MeV of 2NF and 
        2NF + 3NF ($c_E=0.5$).
        The empirical Tjon band is indicated by 
        shaded area, adopted from Ref.~\cite{splatter2005plb}.
        (b) Binding energy correlation between 
        $^3$H and $^{16}$O with notations same as panel~(a). }  
        \label{fig:H3He4O16corr}
    \end{figure*}

    We also examine the well-established correlation of Tjon line  
    between $^3$H and $^4$He~\cite{stjon1975plb}
    by calculations with varying cutoffs.  
    Explicitly, the calculations with 2NF only $(c_E=0)$ and 
    complementary calculations with ``wrong'' value of 3NF
    $(c_E = 0.5$ for all cutoffs) are performed. 
    The results are presented in Fig.~\ref{fig:H3He4O16corr}(a).
    We can find that 2NF-only results and 2NF+3NF ($c_E=0.5$) results 
    are well located within the empirical Tjon band~\cite{splatter2005plb}.
    The 2NF-only and2NF + 3NF ($c_E=0.5$) results 
    are also strongly linearly correlated
    with a Pearson coefficient $r_{xy}=0.999$. The deviation of 
    the correlation line from the experimental value can be attributed
    to the Galilean invariance breaking (GIB) effects. 
    If such effect get restored, as shown in Ref.~\cite{sshi2026plb}, 
    the simultaneous reproduction of both RG invariance an a 
    correlation line passing the experimental value can be achieved.
    We perform analogous calculations to heavier system 
    like $^{16}$O, and observe a resemblance of Tjon line formed by 
    binding energies of $^{16}$O and $^3$H in 
    Fig.~\ref{fig:H3He4O16corr}(b).
    Still both 2NF-only and 2NF + 3NF $(c_E=0.5)$ results are 
    highly correlated with a Pearson coefficient
    $r_{xy}=0.997$ and the experimental value lies just within
    the $1\sigma$ uncertainty band.
    Both cases highlight that a unique value of $c_E$ can be 
    simultaneously determined by binding energies of 
    light nuclei such as $^3$H, $^4$He and 
    medium-mass nuclei like $^{16}$O.
    
    This simple and straightforward finding is contrasted with the 
    recent practices of adapting rather complicated 3NFs to 
    many-body observables.
    We conjecture that the difference is related to the 
    different regulation schemes and need further investigations.
\section*{Brueckner-Hartree-Fock approach}
    The lattice Monte Carlo simulation for nuclear 
    matter using a Hamiltonian with nonzero sign problem
    still remains formidable. However
    traditional approaches, such as the Brueckner-Hatree-Fock (BHF) method, remain highly advantageous.
    Moreover, 3NF can be reduced into density-dependent 
    2NF using normal ordering trick~\cite{shebeler2021pr}, 
    which further reduces the computational consuming.
    In BHF approah, the particle-particle correlations 
    to all orders are incorporated by Brueckner-Goldstone 
    equation~\cite{sday1967rmp}. 
	\begin{equation}\label{Beq}
		G(W)= V + V \frac{Q}{e(W)} G(W).
	\end{equation}
    Here $Q$ denotes the Pauli blocking operator
    \begin{equation}
		Q(\bm{p}_1,\bm{p}_2) = \begin{cases}
			1 & {\rm if}~|\bm{p}_1|>p_F~\&~|\bm{p}_2|>p_F \\
			0 &{\rm otherwise}
		\end{cases},
	\end{equation} 
    with $p_F = (3\pi^2 \rho/2)^{1/3}$ being the Fermi momentum 
    at density $\rho$.
    $e(W) = W - H$ is the energy denominator and $W$ is 
    the starting energy. Given particle-particle state $|\bm{p}_1\bm{p}_2\rangle$, the energy denominator becomes
    \begin{equation}
        e(W) |\bm{p}_1\bm{p}_2\rangle =  W - E(p_1) - E(p_2).
    \end{equation}
    For single-particle energy $E(p)$, the continuous 
    choice~\cite{ssong1998prl} is adopted,
    \begin{equation}
        E(p) = \frac{p^2}{2M_\mathrm{N}} + \sum_{p'}^{p_F}
        G(W=E(p) + E(p')).
    \end{equation}
    After self-consistent solution of single-nucleon 
    energy is achieved, binding energy per nucleon in
    symmetric nuclear matter (SNM) can be 
    computed by
    \begin{equation}
		\frac{E(\rho)}{A}= \frac{3p_F^2}{10M_{\rm N}} +
		\frac{1}{2\rho}\sum_{p_1p_2}^{p_F} G(W = E(p_1) + E(p_2)).
	\end{equation}

    \begin{figure*}
        \centering
        \includegraphics[width=.85\linewidth]
        {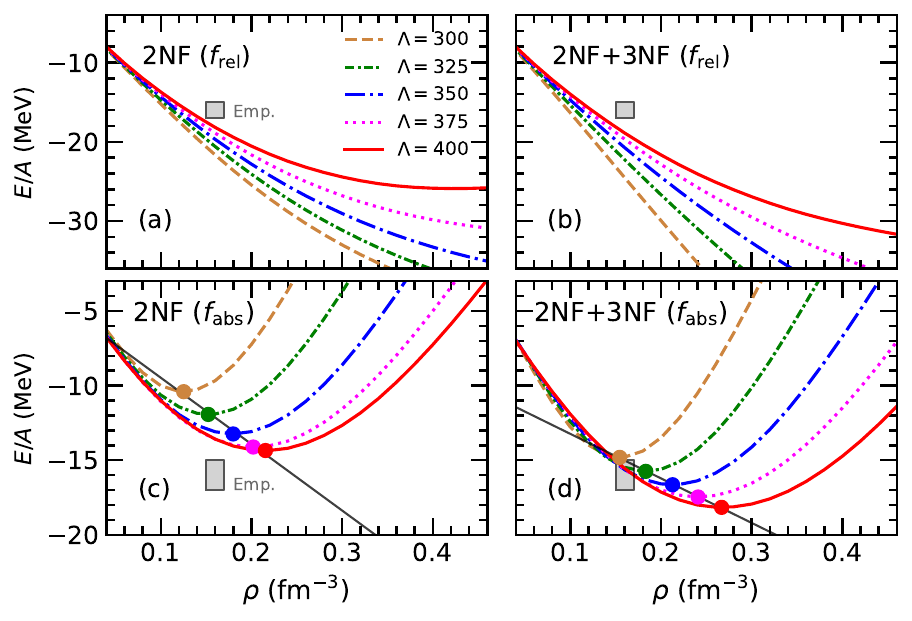}
        \caption{The equations of state in symmetric nuclear matter
        calculated by BHF method using chiral nuclear force 
        with cutoff varying
        from $\Lambda=300$ MeV to $400$.
        (a), (b) Calculations using 2NF and 2NF+3NF respectively with 
        relative-momentum regulator.
        (c), (d) Calculations using 2NF and 2NF+3NF respectively with 
        absolute-momentum regulator with saturation points 
        depicting by dots. Note the scales of $y$-axis are different 
        between upper-two and lower-two panels. The shaded area 
        indicates the empirical saturation region.}
        \label{fig:SNMEOS}
    \end{figure*} 
    In practice, we have employed the angular-averaging 
    approximation and normal ordering for 3NF.
	  The dependence on the total momentum $\bm{P}$ in 
    absolute-momentum regulator $f_{\rm abs}$ is proved to be very 
    weak~\cite{sli2018prc}.
    We can define the angular-averaged single-particle regulator
    \begin{align} \nonumber 
        \bar{f}(P,p)\equiv &\frac{1}{2}\int_{-1}^{+1}\mathrm{d}x~\exp
        \left[-\frac{\left(\bm{P}/2+\bm{p}\right)^6  +
        \left(\bm{P}/2-\bm{p}\right)^6}{2\Lambda^6}\right]   
            \\
        =&e^{-r}\frac{\sqrt{\pi}}{2s}\mathrm{erf}(s).
	\end{align}
	with $x=\hat{P}\cdot\hat{p}$, $r=(P^2/4+p^2)^3/\Lambda^6$,
    $s=Pp \sqrt{3(P^2/4+p^2)}/\Lambda^3$.

    Therefore absolute-momentum regulator for two-nucleon sector 
    can be approximated by  
    \begin{equation}\label{fabs:AA}
    	f_{\rm abs}(p_1',p_1,p_2',p_2) \approx  
        \bar{f}(P,p') \bar{f}(P,p).
	\end{equation}
    The normal-ordered 3NF is obtained by 
	\begin{align} \label{NO3N}
		\langle 1'2'|V_{\overline{\mathrm{3N}}}
        |12\rangle_\mathcal{A} 
		= \sum_{\sigma_3 \tau_3} \int^{p_F}
		\frac{\mathrm{d}\bm{p}_3}{(2\pi)^3} 
		\langle 1'2'3|V_\mathrm{3N}|123\rangle_\mathcal{A},
	\end{align} 
    where $|i\rangle =|\bm{p}_i\sigma_i\tau_i\rangle$,  
    $\sigma_i$, $\tau_i$ are the spin and isospin indices and
    the subscript $\mathcal{A}$ indicates antisymmetrization.

    We use $V=V_\mathrm{2N} + V_{\overline{\mathrm{3N}}}/3$ 
    as the input to Bethe-Brueckner equation~\eqref{Beq}.
    Since $f_{\rm abs}$ is separable, we can derive the 
    non-vanishing normal-ordered 3NF in $^1S_0$ and
    $^3S_1$ channels:
    \begin{equation}\label{NOV3N}
        \langle S|V_{\overline{\mathrm{3N}}}|S\rangle  
        =-\frac{3c_E \rho_\Lambda}{2F_\pi^4 \Lambda_\chi} ~
        f_{\rm abs}(p_1',p_1,p_2',p_2),
    \end{equation}   
    with 	
    \begin{align} \nonumber 
		\rho_\Lambda = & \sum_\sigma\int^{p_F}_0 
        \frac{p^2\mathrm{d} p}{2\pi^2}~\exp
        \left[-\left(\frac{p}{\Lambda}\right)^6\right]
            \\
        &=\frac{\Lambda^3}{6\sqrt{\pi^3}}{\rm erf}
        \left(\frac{p_F^3}{\Lambda^3}\right).
	\end{align} 

    \begin{figure*} 
        \centering
        \includegraphics[width=.75\linewidth]{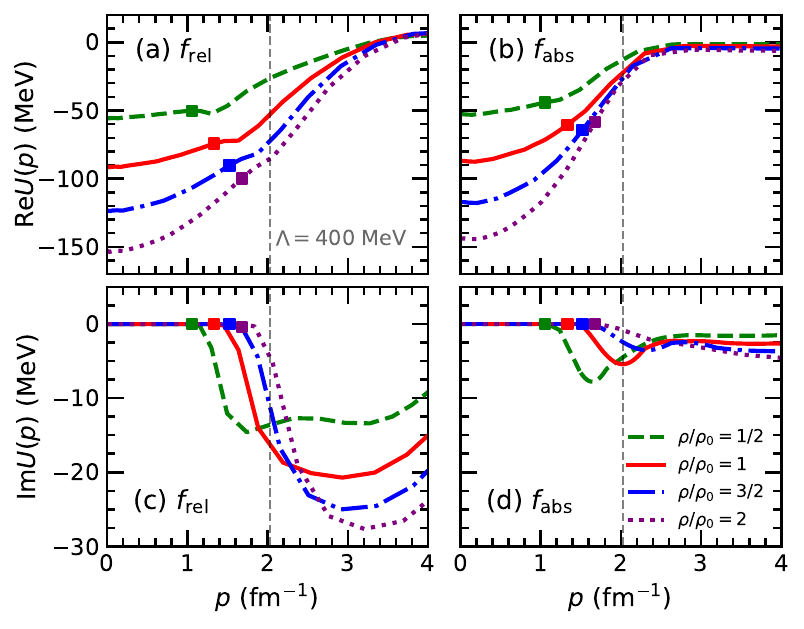}
        \caption{Real and imaginary parts of single-particle potentials
        in symmetric nuclear matter at various densities calculated using 2NF with cutoff $\Lambda=400$~MeV.
        Results with the same LECs and different regulators are contrasted.
        Squares denote the corresponding Fermi momenta.}   
        \label{fig:snm_usp}
    \end{figure*}
    In Fig.~\ref{fig:SNMEOS}, we compare the equation of states
    (EOSs) resulting from different regulator schemes.
    While Fig.~\ref{fig:SNMEOS}(a) already shows considerable overbinding
    for all cutoffs, inclusion of attractive 3NFs in panel~(b) leads
    to further collapse of all EOSs. 
    However, a completely opposite situations are observed when 
    the absolute-momentum regulator is employed, where 2NF-only EOSs
    are underbound by several MeVs and the inclusion of attractive 
    3NFs significantly resolve the discrepancy. 
    We also examine the Coester line (band)
    in SNM~\cite{scoester1970prc} obtained by 2NF only and 2NF + 3NF
    using absolute-momentum regulator. 
    As shown Fig.~\ref{fig:SNMEOS}(c) and (d), the saturation 
    points from 2NF-only or 2NF + 3NF calculations are separately 
    highly correlated.
    Even in 2NF + 3NF calculations, we can still observe a residual 
    cutoff dependence in SNM, unlike the results of finite nuclei.
    This discrepancy can be traced back to the fitting of phase 
    shift up to a relative momentum of $p=200$~MeV. By incorporating
    higher order of chiral force together with  larger range of momentum 
    in fitting, we believe a cutoff-independent results can be 
    obtained in SNM of supra-saturation densities.

    To uncover the microscopic origin of , we compare in Fig.~\ref{fig:snm_usp} the single-particle potentials $U(p)$ generated by both regulators at $\Lambda = 400\;\text{MeV}$.
    While both regulators yield similar results for $p \ll \Lambda$, $f_{\mathrm{abs}}$ suppresses high-momentum attraction far more efficiently.
    In contrast, $f_{\mathrm{rel}}$ exhibits a much slower high-momentum damping, leading to the excessive binding.
    The distinction becomes even more pronounced in the imaginary part, $\text{Im}\,U(p)$, which emerges for $p > p_F$ and quantifies the strength of particle-hole correlations.
    This marked contrast highlights the ability of the \emph{absolute}-momentum regulator to suppress high-momentum contributions, thereby preventing overbinding in both symmetric nuclear matter and medium-mass nuclei.

\section*{Galilean invariance restoration}

We employ a regulator acting on the absolute single-particle momentum, which explicitly breaks Galilean invariance. Such symmetry-breaking effects are common in lattice simulations, as the lattice defines a preferred inertial frame, thereby breaking the equivalence of all inertial frames. In our calculations, we smear the lattice using a continuum single-particle regulator, which restores rotational symmetry~\cite{sklein2015plb}. However, the breaking of Galilean invariance persists.

The consequences of Galilean invariance breaking have been studied within the framework of nuclear lattice effective field theory (NLEFT). A study based on a one-dimensional toy model demonstrated that the inclusion of Galilean invariance restoration (GIR) terms, following Symanzik's improvement program, accelerates the convergence of 
lattice predictions~\cite{sklein2018epja}. 
    In calculations using realistic nuclear forces, GIR terms 
    have been introduced to improve the description of 
    experimental data \cite{sli2019prc,selhatisari2024nat}.

    \begin{table*} 
    \renewcommand{\arraystretch}{1.2}
    \setlength{\tabcolsep}{7.5pt} 
    \centering 
    \caption{The fitted GIR coefficients up to $Q^4$. The 
    three-body force $c_E$ is fitted to triton binding energy E($^3$H)=-8.482 MeV. The results are in lattice unit for lattice spacing $a=(200~{\rm MeV})^{-1}$ . The two-body contact terms and OPEP remain the same as previous 
    section.}
    \begin{tabular}{c|rrrrrrrr}
    \toprule 
    $g_{{}^1S_0}$ & 0.0863 & -0.0314 & -0.0575 & -0.0514 
                  &-0.0358 & -0.0200 & -0.0097
            \\
    $g_{{}^1S_0}'$&-1.173 & -0.532 & -0.262 & -0.141 
                  &-0.0861 & -0.0630 & -0.0570
            \\
    $g_{{}^3S_1}$ & 0.0637 & 0.115 & -0.0533 & -0.0896
                  &-0.0778 & -0.0506 & -0.0235
            \\
    $g_{{}^3S_1}'$&-2.416  & -1.056 & -0.487 & -0.235 
                  &-0.120  & -0.0694 & -0.0535
                    \\
    $c_E~(L=11)$& 1.063 & 0.436 & 0.176 & 0.814 & 0.0690
                  & 0.0969 & 0.167 \\ 
    \bottomrule	 
    \end{tabular}
     \label{tab:GIRQ4LECs}
    \end{table*}   

Recent calculations using the same interactions as in this work demonstrate that including GIR terms significantly improves the renormalization group (RG) invariance of the $^4$He binding energy~\cite{sshi2026plb}. In those studies, we found that calculations without GIR terms overestimated the $^4$He binding energy by approximately $1$~MeV, an error that scales as $\mathcal{O}(\Lambda^{-2})$. This discrepancy is fully resolved by including leading-order GIR terms in the Hamiltonian. It is therefore of interest to examine whether the results presented in this work change upon inclusion of GIR correction terms.

    \begin{figure*}
        \centering
        \includegraphics[width=.75\linewidth]
        {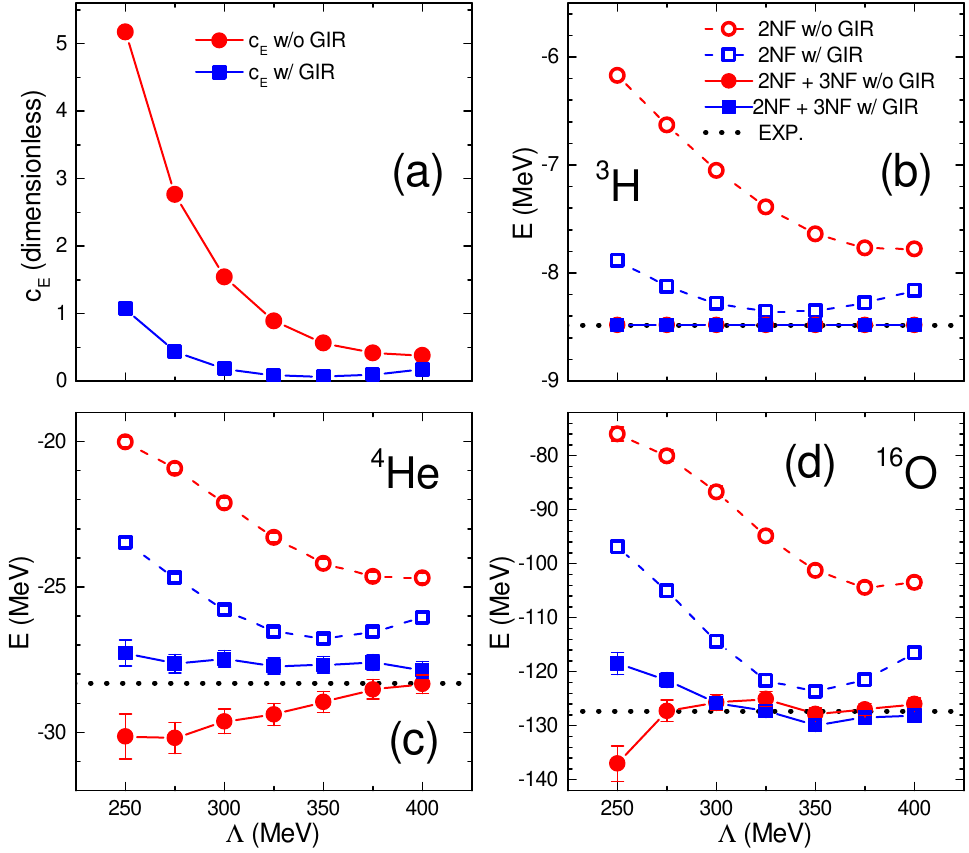}
        \caption{(a) Fitted three-body LEC $c_E$ with and without the Galilean invariance restoration terms.
        (b-d) Binding energies of $^3$H, $^4$He and $^{16}$O calculated with two-body force only (open symbols) and two- plus three-body forces (full symbols). Circles and squares denote the results without and with GIR terms, respectively. Dotted line represents the experimental values. }
        \label{fig:GIRQ4}
    \end{figure*} 

For medium-mass nuclei such as $^{16}$O, we introduce next-to-leading order GIR corrections:
\begin{align*}
V_{\mathrm{GIR}}^{1\mathrm{S}0} &= [g_{1\mathrm{S}0}P^2 + g_{1\mathrm{S}0}^\prime (p^2 + p^{\prime2}) P^2 ] f_{\rm abs}(p_1',p_1, p_2', p_2), \\
V_{\mathrm{GIR}}^{3\mathrm{S}1} &= [g_{3\mathrm{S}1}P^2 + g_{3\mathrm{S}1}^\prime (p^2 + p^{\prime2}) P^2 ] f_{\rm abs}(p_1',p_1, p_2', p_2),
\end{align*}
where $f_{\rm abs}(p_1', p_1, p_2',p_2)$ is the same 
absolute-momentum regulator used elsewhere in this work. The low-energy constants (LECs) $g_{^1S_0/^3S_1}$ and $g_{^1S_0/^3S_1}'$ are determined by requiring that the $S$-wave neutron-proton scattering phase shifts remain independent of the total momentum $\bm{P}$. 
Since nucleons within medium-mass nuclei typically possess larger momenta, we perform a more comprehensive restoration of Galilean invariance up to higher total momenta $|\bm{P}|$. 
For each cutoff value, we refit the three-body LEC $c_E$ to fix the triton binding energy to its experimental value. The running LECs are summarized in Tab.~\ref{tab:GIRQ4LECs}. Note that the two-body LECs remain identical to those used in the rest of this work.

The resulting binding energies for $^3$H, $^4$He, and $^{16}$O are shown in Fig.~\ref{fig:GIRQ4}. We have employed recently developed techniques to incorporate the complex GIR terms \cite{sliu2025epja}. We find that the results including both GIR terms and three-body forces maintain excellent RG invariance within the interval $300$~MeV $\le \Lambda \le 400$~MeV, consistent with our expectations. The agreement with experimental data is also remarkable. The slight deviation in $^4$He is primarily due to charge-symmetry-breaking effects, which were omitted here for simplicity. Notably, when these terms were included in Ref.~\cite{sshi2026plb}, the resulting $^4$He binding energy agreed with experiment to within $100$~keV.

\setcounter{enumiv}{0}
%

\end{document}